\documentclass[12pt]{article}

\usepackage{amssymb}
\usepackage{amsmath}
\usepackage{amscd}
\usepackage{latexsym}
\usepackage{graphicx}

\usepackage{cite}

\newcommand{\be}{\begin{equation}}
\newcommand{\ee}{\end{equation}}
\newcommand{\Dlt}{\Delta}
\newcommand{\dlt}{\delta}
\newcommand{\prt}{\partial}
\newcommand{\br}{{\bf r}}
\newcommand{\bk}{{\bf k}}
\newcommand{\bn}{{\bf n}}

\newcommand{\bfe}{{\bf e}}
\newcommand{\bd}{{\bf d}}
\newcommand{\bP}{{\bf P}}

\newcommand{\bt}{\beta}

\newcommand{\ep}{\varepsilon}

\newcommand{\ra}{\rightarrow}
\newcommand{\sgm}{\sigma}
\newcommand{\Sgm}{\Sigma}

\newcommand{\om}{\omega}

\newcommand{\Gm}{\Gamma}
\newcommand{\dgr}{\dagger}

\newcommand{\Lbd}{\Lambda}

\newcommand{\rgl}{\rangle}
\newcommand{\lgl}{\langle}

\begin{document}

\begin{center}

{\Large{\bf Statistical systems with nonintegrable interaction potentials} \\ [5mm]

V.I. Yukalov} 

{\it
Bogolubov Laboratory of Theoretical Physics, \\
Joint Institute for Nuclear Research, Dubna 141980, Russia }

\end{center}

\vskip 5cm

\begin{abstract}

Statistical systems composed of atoms interacting with each other trough nonintegrable
interaction potentials are considered. Examples of these potentials are hard-core 
potentials and long-range potentials, for instance, the Lennard-Jones and dipolar 
potentials. The treatment of such potentials is known to confront several problems,
e.g., the impossibility of using the standard mean-field approximations, such as Hartree
and Hartree-Fock approximations, the impossibility of directly introducing coherent 
states, the difficulty in breaking the global gauge symmetry, which is required for
describing Bose-Einstein condensed and superfluid systems, the absence of a correctly 
defined Fourier transform, which hampers the description of uniform matter as well 
as the use of local-density approximation for nonuniform systems. A novel iterative 
procedure for describing such systems is developed, starting from a correlated 
mean-field approximation, allowing for a systematic derivation of higher orders, and 
meeting no problems listed above. The procedure is applicable to arbitrary systems,
whether equilibrium or nonequilibrium. The specification for equilibrium systems is
presented. The method of extrapolating the expressions for observable quantities 
from weak coupling to strong coupling is described.      

\end{abstract}

\vskip 2mm

{\bf PACS numbers}: 05.30.Ch 

\vskip 2mm

\newpage

\section{Introduction}

Atomic interactions are usually described by pair interaction potentials. Quite 
often, such potentials are not integrable. This essentially complicates the use 
of these potentials for developing the description of statistical systems. Thus, 
nonintegrable interaction potentials do not allow for the use of the standard 
mean-field approximations, such as Hartree, Hartree-Fock, or Hartree-Fock-Bogolubov 
approximations. For treating the systems with these potentials, one needs to resort 
to two-particle characteristics solving the Brueckner or Bethe-Salpeter equations 
(see, e.g., \cite{Thouless_1,Manassah_2,Greiner_3}). Dealing with two-particle 
theories is more complicated than with mean-field approximations and, in addition,
it is not always clear how to develop a procedure for obtaining higher-order 
consecutive corrections above the given two-particle approximation. It would 
certainly be desirable to have a method of successive iterations that would 
combine the simplicity of using, as a first step, a mean-field approximation and 
confronting no divergences related to the nonintegrable interaction potential. 

Among other principal difficulties arising when dealing with nonintegrable potentials,
it is possible to mention the impossibility of introducing coherent states, the 
problem with breaking the global gauge symmetry required for characterizing systems 
with Bose-Einstein condensate and superfluid systems, and the absence of well defined 
Fourier transforms, which are necessary for describing uniform systems, as well as 
nonuniform systems in the local-density approximation. The explicit illustration of 
these difficulties will be given in the following section. 

The aim of the present paper is to suggest a consistent iterative procedure allowing 
for the possibility of starting with a mean-field-type approximation containing no 
divergences and providing an explicit method for deriving consecutive higher-order 
approximations. For the sake of generality, the iterative procedure is formulated in
the language of Green functions, so that its application can be realized for arbitrary 
systems with nonintegrable potentials, whether equilibrium or not. If the interaction 
potential is integrable, the suggested iterative approach reduces to the standard 
iteration theory for Green functions. As examples of nonintegrable potentials, the 
Lennard-Jones and dipolar potentials are considered, showing the possible way of 
regularizing them. 

The general procedure is applicable to arbitrary systems, whether equilibrium or not.
The specification for equilibrium systems is considered. General rules for defining 
smoothing functions regularizing interaction potentials are given. A method is 
described, based on self-similar approximation theory, allowing for the extrapolation 
of the values of observable quantities from the region of weak coupling to arbitrarily 
strong coupling. Using this method, it is possible to derive a rather simple expression
for the ground-state energy of a Bose gas with hard-core interactions, which is in
very good agreement with Monte Carlo simulations.   

Throughout the paper the system of units is used where the Planck and Boltzmann 
constants are set to one.

\section{Problems with nonintegrable interaction potentials}

Let us denote by $x$ the set of the spatial variables ${\bf r}$ and of internal 
degrees of freedom, such as spin, if any. Employing field operators, we shall omit, 
when there can be no ambiguity, the notation of time $t$, writing $\psi(x)$ instead 
of $\psi(x,t)$ and restoring time, when it is important. Depending on statistics, 
the field operators satisfy either Bose or Fermi commutation relations,
\be
\label{1}
\left [ \psi(x) , \; \psi^\dgr(x') \right ]_\mp = \dlt(x-x') \; , \qquad
\left [\psi(x) , \; \psi(x') \right ]_\mp = 0 \; .
\ee
The system Hamiltonian has the form
\be
\label{2}
 H = \int \psi^\dgr(x) [ K(x) - \mu(x) ] \psi(x)\; dx + \frac{1}{2}
\int \psi^\dgr(x) \psi^\dgr(x') V(x,x') \psi(x')\psi(x) \; dx dx' \; ,
\ee
in which 
$$
K(x) = -\; \frac{\nabla^2}{2m} + U(x) \;   ,
$$
$U(x)$ is an external potential, $V(x,x^\prime) = V(x^\prime,x)$ is the interaction 
potential, and $\mu(x)$ is a local chemical potential including external 
perturbing fields. 

The interaction potential is assumed to be nonintegrable, such that
\be
\label{3}
  \left | \int  V(x,x') \; dx' \right | ~\ra ~ \infty \;  .
\ee
It is this divergence that leads to difficulties, as is explained below. 

\vskip 3mm

{\bf A. Mean-field approximation}

\vskip 2mm

Nonintegrability (\ref{3}) yields divergences when resorting to the standard 
mean-field approximations. For instance, let us consider the Hartree-Fock 
approximation
$$
\psi_1^\dgr \psi_2^\dgr \psi_2 \psi_1 = 
\lgl \psi_1^\dgr \psi_1 \rgl \psi_2^\dgr \psi_2 +
$$
\be
\label{4}            
 +
\psi_1^\dgr \psi_1 \lgl \psi_2^\dgr \psi_2 \rgl \pm
\lgl \psi_1^\dgr \psi_2 \rgl \psi_2^\dgr \psi_1 \pm
\psi_1^\dgr \psi_2 \lgl \psi_2^\dgr \psi_1 \rgl -
\lgl \psi_1^\dgr \psi_1 \rgl \lgl \psi_2^\dgr \psi_2 \rgl \mp
\lgl \psi_1^\dgr \psi_2 \rgl \lgl \psi_2^\dgr \psi_1 \rgl \; ,
\ee
where, for short, we denote $\psi_i \equiv \psi(x_i)$. Substituting this into
the Hamiltonian results in the generally divergent Hartree potential
\be
\label{5}
\left | \int  V(x,x') \rho(x') \; dx' \right | ~\ra ~ \infty \;   ,
\ee
where the density is given by the statistical average
$$
 \rho(x) \equiv \lgl \psi^\dgr(x) \psi(x) \rgl \;  .
$$
The divergence becomes evident in the uniform case, when the density is constant. 

Hence, the mean-field approximation cannot be used for a nonintegrable interaction 
potential.

\vskip 2mm

{\bf B. Coherent states}

\vskip 2mm

Coherent state is defined as an eigenstate of the destruction field operator,
\be
\label{6}
 \psi(x,t) | \eta \rgl = \eta(x,t) | \eta \rgl \;   ,
\ee
with the eigenvalue called coherent field. Then the equation of motion for the 
field operator 
\be
\label{7}
i\; \frac{\prt}{\prt t} \; \psi(x,t) = \frac{\dlt H}{\dlt\psi^\dgr(x,t) } 
\ee
results in the nonlinear Schr\"{o}dinger equation for the coherent field
\be
\label{8}
  i\; \frac{\prt}{\prt t} \; \eta(x,t) = \left [ K(x) - \mu(x) + 
\int V(x,x') | \eta(x',t) |^2 \; dx' \right ] \; \eta(x,t) \; .
\ee
If the interaction potential is nonintegrable, then, in general, the integral term
in the right-hand side diverges:
\be
\label{9}  
  \left | \int V(x,x') | \eta(x',t) |^2 \; dx' \right | ~ \ra ~ \infty \; .
\ee
Again, the divergence is evident for a uniform case, when the coherent field is 
constant. 

This implies that the usual way of introducing coherent states does not work,
when the interaction potential is not integrable.  

\vskip 2mm

{\bf C. Bose-Einstein condensation}

\vskip 2mm

For the phenomenon of Bose-Einstein condensation, as is known \cite{Lieb_4,Yukalov_5},
the global gauge symmetry breaking is a necessary and sufficient condition. The 
symmetry breaking can be accomplished in several equivalent ways, the simplest of 
which is by means of the Bogolubov \cite{Bogolubov_6,Bogolubov_7} shift
\be
\label{10}
 \psi(x) = \eta(x) + \psi_1(x) \;  ,
\ee
in which the first term is the condensate function and the second, an operator of
uncondensed atoms. Note that this is an exact canonical transformation 
\cite{Yukalov_5,Yukalov_41,Yukalov_8}, but not an approximation as sometimes is 
assumed. The Bogolubov shift defines the condensate function as an order parameter
\be
\label{11}
  \eta(x) =  \lgl \psi(x) \rgl \; .
\ee
The equation of motion for the condensate function
\be
\label{12}
  i\; \frac{\prt}{\prt t} \; \eta(x,t) = 
\left \lgl \frac{\dlt H}{\dlt\eta^*(x,t)} \right \rgl
\ee
takes the form
$$
i\; \frac{\prt}{\prt t} \; \eta(x,t) = [ K(x) - \mu_0(x) ] \eta(x) + 
$$
\be
\label{13}
 +
\int V(x,x') \left [ \rho(x') \eta(x) + \rho_1(x,x') \eta(x') +
\sgm_1(x,x') \eta^*(x') + \xi(x,x') \right ] \; dx' \;  ,
\ee
where the notations are used for the single-particle density matrix
$$
  \rho_1(x,x') \equiv \lgl \psi_1^\dgr(x') \psi_1(x) \rgl \;  , 
$$
anomalous averages
$$
  \sgm_1(x,x') \equiv \lgl \psi_1(x') \psi_1(x) \rgl \;  , \qquad
 \xi(x,x') \equiv \lgl \psi_1^\dgr(x') \psi_1(x') \psi_1(x) \rgl \; ,
$$
and the total density
$$
\rho(x) = \rho_0(x) + \rho_1(x) \;   ,
$$
consisting of the condensate density
$$
\rho_0(x) \equiv | \eta(x) |^2 \; ,
$$
and the density of uncondensed atoms
$$
 \rho_1(x) = \rho_1(x,x) = \lgl \psi_1^\dgr(x) \psi_1(x) \rgl  \; .
$$
Equation (\ref{13}) contains the Hartree term (\ref{5}), that is generally divergent.       

Thus, the global gauge symmetry breaking, that is required for a correct description 
of Bose-condensed systems, cannot be realized. For instance, the symmetry breaking 
leads to the appearance of anomalous averages that need to be accurately calculated 
\cite{Yukalov_42} for obtaining the condensate fraction in agreement with numerical 
data \cite{Rossi_43}.

\vskip 2mm

{\bf D. Superfluid state}

\vskip 2mm

The analogous problem arises when considering superfluid systems in three-dimensional space, 
since superfluidity is accompanied by Bose-Einstein condensation, which requires global gauge 
symmetry breaking. Then, breaking the symmetry by the Bogolubov shift (\ref{10}), we again
get a divergent term in the condensate-function equation. 

It is easy to show that without gauge symmetry breaking, superfluidity in three-dimensional
systems cannot be defined. The general formula for the superfluid density reads as
\be
\label{14}
 \rho_s = \rho - \; \frac{2Q}{3T} \;  ,
\ee
where the dissipated heat  
\be
\label{15}
Q = \frac{{\rm var}(\hat\bP)}{2m N}
\ee
is expressed through the variance
$$
{\rm var}(\hat\bP) = \lgl \hat\bP^2 \rgl - \lgl \hat\bP \rgl^2
$$
of the momentum operator
$$
  \hat \bP = \int \psi^\dgr(\br) ( - i\vec{\nabla} ) \psi(\br)\; d\br \; .
$$

The dissipated-heat expression contains the anomalous averages that cannot be omitted.
Thus, in the Hartree-Fock-Bogolubov approximation, the dissipated heat is
\be
\label{16}
 Q = 
\int \frac{k^2}{2m} \left ( n_k + n_k^2 - \sgm_k^2 \right ) \frac{d\bk}{(2\pi)^3} \;  ,
\ee
where
$$
n_k = \int \rho_1(\br,0) e^{-i\bk\cdot\br} \; d\br \; , \qquad
 \sgm_k = \int \sgm_1(\br,0) e^{-i\bk\cdot\br} \; d\br \;  .
$$
By direct calculations \cite{Yukalov_5,Yukalov_44,Yukalov_45} it is straightforward 
to prove that omitting the anomalous average $\sigma_k$ results in the divergence 
of integral (\ref{16}).  

In this way, breaking the global gauge symmetry, which is necessary for the correct 
description of superfluid systems in three dimensions, leads to divergences, similar 
to those occurring in the case of Bose-condensed systems. 

\vskip 2mm

{\bf E. Fourier transform}

\vskip 2mm

In the case of a uniform system or employing the local-density approximation for a
nonuniform system, one needs to consider the Fourier transform of the interaction 
potential. For instance, keeping in mind the potential
\be
\label{17}
V(\br,\br') = V(\br - \br') \; ,
\ee
one considers the Fourier transform
\be
\label{18}
 V_k = \int V(\br) e^{-i\bk\cdot\br} \; d\br \;  ,
\ee
with the inverse transform
\be
\label{19}
  V(\br) = \frac{1}{V} \sum_k V_k e^{i\bk\cdot\br} \; .
\ee
But if the potential is not absolutely integrable, such that
\be
\label{20}
 \int | V(\br) | \; d\br ~ \ra ~ \infty \;  ,
\ee
then the Fourier transform $V_k$ is not well defined \cite{Champeney_46,Yukalov_47}. 
And if the interaction potential is not integrable, it is not absolutely 
integrable, since
$$
 \left | \int  V(\br)  \; d\br \right | \leq \int | V(\br) | \; d\br \;  .
$$

In the following sections, we develop an iterative procedure that is free from all
those problems discussed above, despite that the interaction potential is not 
integrable.

\section{Iterative procedure for Green functions}

To make formulas more compact, let us introduce the following abbreviated notations
for functions
\be
\label{21}
 f(12\ldots n) \equiv f(x_1,t_1,x_2,t_2,\ldots,x_n,t_n) \;  ,
\ee
e.g., for the delta function
\be
\label{22}
 \dlt(12) = \dlt(x_1 - x_2) \dlt(t_1 - t_2) \;  ,
\ee
and for differentials
\be
\label{23}
 d(12\ldots n) \equiv dx_1 dt_1 dx_2 dt_2 \ldots dx_n dt_n \;  .
\ee
And let us define the interaction potential
\be
\label{24}
 V(12) \equiv  V(x_1 - x_2) \dlt(t_1 - t_2 + 0) \; .
\ee

The single-particle Green function, or propagator, reads as
\be
\label{25}
 G(12) = - i \lgl \hat T \psi(1) \psi^\dgr(2) \rgl \;  ,
\ee
with $\hat T$ being the chronological operator. For coinciding arguments, 
one has
\be
\label{26}
 G(11) \equiv \lim_{x_2\ra x_1} \; \lim_{t_2\ra t_1+0} G(12) \;  ,
\ee
which defines the particle density
\be
\label{27}
 \rho(1) = \pm i G(11) \;  .
\ee
The two-particle Green function is
\be
\label{28}
 G_2(1234) = - \lgl \hat T \psi(1) \psi(2) \psi^\dgr(3) \psi^\dgr(4) \rgl \;  .
\ee

Introducing the inverse propagator
\be
\label{29}
 G^{-1}(12) = \left [ i \; \frac{\prt}{\prt t_1} - K(1)  + \mu(1) \right ] \dlt(12)
-\Sigma(12) \;  ,
\ee
with the self-energy
\be
\label{30}
\Sigma(12) = \pm i \int V(13) G_2(1334) G^{-1}(42)\; d(34) \;   ,
\ee
the equation of motion for the single-particle propagator can be written as
\be
\label{31}
 \int  G^{-1}(13) G(32)\; d(3) = \dlt(12) \;  .
\ee

Choosing a convenient zero approximation for the inverse propagator
\be
\label{32}
 G_0^{-1}(12) = \left [ i \; \frac{\prt}{\prt t_1} - K(1) + 
\mu(1) \right ] \dlt(12) - \Sigma_0(12) \;  ,
\ee
with the related equation of motion
\be
\label{33}
 \int G^{-1}_0(13) G_0(32)\; d(3) = \dlt(12) \; ,
\ee
one gets the Dyson equation
\be
\label{34}
 G(12) =  G_0(12) + \int G_0(13)  \left [\; \Sigma(34) - \Sigma_0(34)\; \right ] G(42) \; d(34) \; .
\ee
The latter is to be solved by the iterative procedure
\be
\label{35}
 G_n ~ \ra ~ \Sigma_{n+1} ~ \ra ~ G_{n+1} \;  ,
\ee
which shows that we need the sequence of approximations for the self-energy.

\section{Iterative procedure for self-energy}

As is known and has been explained above, the standard perturbation theory for 
self-energy leads to divergent terms, when the interaction potential is not 
integrable. Here we suggest an iterative procedure for self-energy containing 
no divergences. 

Recall that the two-particle propagator can be written \cite{Kadanoff_9} in the 
Schwinger representation as
\be
\label{36}
 G_2(1223) = G(13) G(22) \mp \frac{\dlt G(13)}{\dlt\mu(2)} \;  .
\ee
Varying the equation of motion (\ref{31}) yields the equation for the two-particle 
propagator,
$$
 G_2(1223) = G(13) G(22) \pm G(12) G(23) +
$$
\be
\label{37}
 + 
\int G(14) G(53) \; \frac{\dlt\Sigma(45)}{\dlt G(67)} \; \left [\; G_2(6227) -
G(67) G(22)\; \right ] \; d(4567) \; .
\ee

Let us introduce a function $D(123)$ by the relation
\be
\label{38}
 s(12) D(123) = \int G_2(1224) G^{-1}(43) \; d(4) \;  ,
\ee
in which the correlation function $s(12)$ will be specified later. Then the 
two-particle propagator becomes
\be
\label{39}
 G_2(1223) \equiv s(12) \int D(124) G(43)\; d(4) \;  .
\ee
The latter equation shows that, by means of the function $D(123)$, the 
single-particle propagator is transformed into the two-particle propagator. 
Therefore $D(123)$ can be called the doubling function. 

In that way, self-energy (\ref{30}) now reads as
\be
\label{40}
\Sigma(12) = \pm i \int \Phi(13) D(132) \; d(3) \;   ,
\ee
where we define the effective potential
\be
\label{41}
 \Phi(12) = s(12) V(12) \;  .
\ee
The function $s(12)$ has to be chosen such that the effective potential be integrable,
\be
\label{42}
 \left | \int \Phi(12) \; d(2) \right | < \infty \;  ,
\ee
because of which the function $s(12)$ can be called smoothing function. 

From Eqs. (\ref{37}) and (\ref{38}), we obtain the equation for the doubling function
$$
s(12) D(123) = D_0(123) +
$$
\be
\label{43}
  + \int G(14) \; \frac{\dlt\Sigma(43)}{\dlt G(56)} 
\left [ s(52) \int D(527) G(76) \; d(7) - G(56) G(22) \right ] \; d(456) \;  ,
\ee
where 
\be
\label{44}
D_0(123) \equiv \dlt(13) G(22) \pm G(12) \dlt(23) \;  .
\ee

It is important to notice that the use of form (\ref{44}) in Eq. (\ref{39}) 
results in the two-particle propagator  
\be
\label{45}
G_2^0(1223) \equiv s(12) \int D_0(124) G(43)\; d(4) \;  ,
\ee
which takes into account the correlation function $s(12)$, being
\be
\label{46}
G_2^0(1223) = s(12) \left [G(13) G(22) \pm G(12) G(23) \right ] \; .
\ee 
This always leads to the occurrence of the effective potential (\ref{41}),
so that no divergences arise.

Let us introduce an operator $\hat{X} = \hat{X}[\Sigma]$, whose action on a 
function $f(123)$ is defined by the equation
$$
\hat X f(123) = [\; 1 - s(12)\; ] f(123) +
$$
\be
\label{47}
 + \int G(14) \frac{\dlt\Sigma(43)}{\dlt G(56)} 
\left [ s(52) \int f(527) G(76)\;  d(7) - G(56) G(22) \right ] \; d(456) \; .
\ee
Then Eq. (\ref{43}) takes the form
\be
\label{48}
 ( 1 - \hat X) D(123) = D_0(123) \;  .
\ee
The latter can be rewritten as
\be
\label{49}
  D(123) = ( 1 - \hat X)^{-1} D_0(123) \; .
\ee
Here, the inverse function of an operator expression is defined in the usual way 
through the expansion
\be
\label{50}
  ( 1 - \hat X)^{-1} = \sum_{n=0}^\infty \hat X^n \; .
\ee
The other representation of the inverse operator function is
\be
\label{51}
( 1 - \hat X)^{-1} = \lim_{n\ra\infty}  \hat Y_n \; ,
\ee
where
\be
\label{52}
\hat Y_n = \sum_{m=0}^n \hat X^m  \qquad (\hat Y_0 = \hat 1)  \;   .
\ee
Then Eq. (\ref{49}) defines the sequence of iterative approximations for the 
doubling function
\be
\label{53}
 D_n(123) = \hat Y_n D_0(123) \;  .
\ee
As a result, we come to the iterative procedure
\be
\label{54}
 D_n ~ \ra ~ \Sigma_{n+1} ~ \ra ~ \hat Y_{n+1} ~ \ra ~ D_{n+1} \; .
\ee

To illustrate this iterative procedure, let us start with the zero-order 
approximation for the doubling function (\ref{44}), substituting which into 
Eq. (\ref{40}), we get the first-order self-energy
\be
\label{55}
 \Sigma_1(12) = \dlt(12) \int \Phi(13) \rho(3) \; d(3) + i \Phi(12) G(12) \;  .
\ee
Using the operator 
\be
\label{56}
\hat Y_1 = 1 + \hat X[\Sigma_1] 
\ee
in Eq. (\ref{53}) yields the first-order doubling function
$$
D_1(123) = D_0(123) [\; 2 - s(12)\;] \pm 
 i \int V(1234) G(42) \; d(4) \mp
$$
\be
\label{57}
\mp i \int V(1443) G(22) [\; 1 - s(43) \; ] \; d(4) \; ,
\ee
with the vertex  
$$
V(1234) = G(14) G(23) \Phi(43) \pm G(13) G(24) \Phi(34) \; .
$$

Employing $D_1(123)$ in Eq. (\ref{40}) results in the second-order approximation
for the self-energy
\be
\label{58}
 \Sigma_2(12) = \Sigma_1(12) + \Dlt(12) + \Lambda(12) \; ,
\ee
in which the correcting term is
$$
\Dlt(12) = \dlt(12)  \int \Phi(13) \rho(3) [\; 1 - s(13)\; ] \; d(3) +
i \Phi(12) G(12) [\; 1 - s(12) \;] \; +
$$
\be
\label{59}
 + \int \Phi(14) G(44) [\; 1 - s(34)\; ] V(1332)\; d(34)  \;  ,
\ee
and the last term is
\be
\label{60}
\Lambda(12) = - \int \Phi(13) G(43) V(1324) \; d(34) \; .
\ee
In that way, the iterative procedure can be continued to any desired order.

First of all, we see that nowhere there appears the divergent bare interaction 
potential, but everywhere we meet only the smoothed effective potential that is 
integrable according to Eq. (\ref{42}). Hence, no divergences occur in the 
iterative process.

Moreover, the smoothing function $s(12)$ can be specified so that to simplify 
the resulting expressions. Thus, if $s(12)$ is chosen to represent a screening 
function, then it enjoys the following properties. When the bare interaction 
provokes divergences, then $s \ra 0$, while when the bare interactions
are finite, then $s \ra 1$. So that in any case the product $s(1-s)$ is small.
If so, then the correction $\Delta$ is small as compared to $\Sigma_1$. As is 
evident, all expressions can be considered as expansions in powers of $\Phi$ and
$1 - s$. Therefore, the last term in correction (\ref{59}) is of third order
and should be omitted in the second-order approximation.  

If in the second-order self-energy (\ref{58}) we neglect the small correcting 
term (\ref{59}), then the self-energy equals 
$$
 \Sigma_2(12) = \Sigma_1(12) - \int \Phi(13) G(43) V(1324) \; d(34) \;  .
$$
But the latter form is the same as would be the second-order approximation for the 
Hamiltonian, in which from the very beginning we would take the effective potential
$\Phi(12)$, instead of the bare potential $V(12)$, that is, if we would accept the
Hamiltonian
$$
 H = \int \psi^\dgr(x) [ K(x) - \mu(x) ] \psi(x) \; dx +
\frac{1}{2} \int \psi^\dgr(x) \psi^\dgr(x') \Phi(x,x') \psi(x')\psi(x) \; dx dx' \; ,
$$
instead of that given by Eq. (\ref{2}). The iterative procedure for both these 
Hamiltonians yields the same first-order self-energy. In higher orders $n > 1$, 
the difference between the iterative terms for these Hamiltonians is characterized
by corrections of the type $\Delta$ that, because of the structure of the operator 
$\hat{X}$, defined in Eq. (\ref{47}), always contain the product $s(1-s)$. Choosing 
the smoothing function as a screening function, such that $s(1-s)$ be small, makes 
small the difference between the terms of the iterative procedure with the bare 
potential and with the effective potential.     

Therefore, if the bare interaction potential is not integrable, it is possible 
to replace it by an effective interaction potential that is integrable and does 
not lead to divergences. Appropriately choosing the smoothing function makes 
the difference in the sequence of the approximations for the iterative procedure
with bare and effective potentials small.

\section{Iterative procedure for response functions}

Different response functions characterize collective properties of statistical 
systems. For example, the response function
\be
\label{61}
\chi(12) \equiv - \; \frac{\dlt\rho(1)}{\dlt\mu(2)} 
\ee
describes collective excitations, with its poles defining the spectrum of collective 
excitations. Having the sequence of approximations for the self-energy makes it 
straightforward to derive the related sequence of approximations for the response 
function.

It is useful to introduce the three-point response function
\be
\label{62}
 \chi(123) \equiv \mp i \; \frac{\dlt G(12)}{\dlt\mu(3)} \;  ,
\ee
whose particular form gives the response function (\ref{61}) as
\be
\label{63}
 \chi(12) = \chi(112) \;  .
\ee
Invoking the Schwinger representation (\ref{36}) gives
\be
\label{64}
 \chi(123) = i [ \; G_2(1332) - G(12) G(33) \; ] \; , \qquad
  \chi(12) = i [ \; G_2(1221) - G(11) G(22) \; ] \;.
\ee
Because of the symmetry property of the two-particle propagator
\be
\label{65}
 G(1221) = G(2112) \;  ,
\ee
the response function (\ref{61}) is symmetric:
\be
\label{66}
 \chi(12) = \chi(21) \;  .
\ee

Introducing the notation
\be
\label{67}
 \chi_0(123) = \pm i G(13)G(32) \; , \qquad \chi_0(12) = \pm i G(12)G(21) \;  ,
\ee
and using Eq. (\ref{37}), we obtain the equation for the response function
\be
\label{68}
 \chi(123) = \chi_0(123) + 
\int G(14) G(52) \; \frac{\dlt\Sigma(45)}{\dlt G(57)} \; \chi(673) \; d(4567) \;  .
\ee
From here, it is clear that the sequence of approximations for the response function
is prescribed by the sequence of the self-energies:
\be
\label{69}
 \Sigma_n ~ \ra ~ \chi_{n+1} \;  .
\ee

Thus, taking for the zero-order self-energy the Hartree expression
\be
\label{70}
\Sigma_0(12) = \dlt(12) \int \Phi(13)\rho(3) \; d(3)
\ee
leads to the equation
\be
\label{71}
 \chi_1(123) = \chi_0(123) + \int \chi_0(124) \Phi(45) \chi_1(553) \; d(45) \;  .
\ee
Respectively, the response function (\ref{61}) is defined by the equation
\be
\label{72}
\chi_1(12) = \chi_0(12) + \int \chi_0(13) \Phi(34) \chi_1(42) \; d(34) \;   .
\ee
The solution to the latter has the form
\be
\label{73}
 \chi_1 = \frac{\chi_0}{1-\chi_0\Phi} \;  ,
\ee
in which one recognizes the random-phase approximation, however with the integrable 
effective potential instead of the nonintegrable bare potential. Taking for the 
self-energy the first-order approximation (\ref{55}) produces the equation
\be
\label{74}
 \chi_2(123) = \chi_0(123) + 
\int [\; \chi_0(124) \chi_2(553) + i G(14) G(52)\chi_2(453) \; ] \Phi(45) \; d(45)\; ,
\ee
from which it follows the equation for the response function (\ref{61}),
\be
\label{75}
\chi_2(12) = \chi_0(12) + 
\int [\; \chi_0(13) \chi_2(42) \pm \chi_0(431) \chi_2(342) \; ] \Phi(34) \; d(34)\; .
\ee

Since in all orders only the effective potential enters the equations, no 
divergences arise.

\section{Examples of nonintegrable interaction potentials}

Depending on the type of the nonintegrable interaction potential, different smoothing
functions can be employed \cite{Yukalov_39}.

\vskip 2mm

{\bf A. Hard-core potentials}

\vskip 2mm

A hard-core potential diverges, when the distance $r \equiv |\br|$ is shorter 
than a hard-core radius $\sigma$, for $r \leq \sigma$, and is finite for larger
distances. For such potentials one uses the simple smoothing function 
\begin{eqnarray}
\label{76}
s(r) = \left \{ \begin{array}{ll}
0 , ~ & ~r \leq \sgm \\
1 , ~ & ~r > \sgm \; ,
\end{array} \right.
\end{eqnarray}
which is called the cutoff regularization. 

A more elaborate smoothing function can be taken in the form
\be
\label{77}
 s(r) = \exp\{ - \bt V(r) \} \;  ,
\ee
where, generally, $\beta$ is a positive parameter. At high temperatures $\beta$ 
can be accepted as inverse temperature $1/T$, while at low temperatures, it is 
to be proportional to the inverse average kinetic energy that is finite even at 
zero temperature due to quantum fluctuations. 

\vskip 2mm

{\bf B. Lennard - Jones potential}

\vskip 2mm

The popular Lennard - Jones potential is
\be
\label{78}
V(r) = 4\ep \left [ \left ( \frac{\sgm}{r}\right )^{12} -  
\left ( \frac{\sgm}{r}\right )^6 \right ] \;  .
\ee
It has a minimum $V(r_0) = - \varepsilon$ at $r_0 = 2^{1/6} \sigma$. 

The smoothing function can be defined as the modulus squared of the radial wave 
function satisfying the zero-energy Schr\"{o}dinger equation \cite{Buendia_10}.
In the quasiclassical approximation, this leads \cite{Yukalov_11} to 
\be
\label{79}
s(r) = \exp \left \{ - b_0  \left ( \frac{\sgm}{r}\right )^5 \right \} \;    ,
\ee
where 
$$
 b_0 \equiv \frac{4}{5\Lbd} \; , \qquad 
\Lbd \equiv \frac{1}{\sqrt{m\ep\sgm^2}} \;  .
$$
Here $\Lambda$ is the de Boer parameter. For instance, in the case of $^4$He,
the Lennard - Jones parameters \cite{Sciver_12} are $\varepsilon = 10.22 K$ 
and $\sigma = 2.556$ \AA, which gives $\Lambda = 0.43$ and $b_0 = 1.86$.  
  
\vskip 2mm

{\bf C. Dipolar potential}

\vskip 2mm

There are numerous statistical systems consisting of particles interacting through 
dipolar forces, for instance, many atomic and molecular gases \cite{Baranov_13}, 
polymers \cite{Barford_14}, biological solutions \cite{Cameretti_15,Waigh_16}, 
and various materials composed of magnetic nanomolecules and nanoclusters 
\cite{Kahn_17,Barbara_18,Kodama_19,Hadjipanays_20,Wernsdorfer_21,Yukalov_22, 
Ferre_23,Yukalov_24,Yukalov_25,Bedanta_26,Berry_27,Beveridge_28}.
The dipolar potential, describing the interaction between two dipoles at distance
$r$ from each other, is
\be
\label{80}
D(\br) = \frac{1}{r^3} \left [ (\bd_1 \cdot\bd_2 ) - 
3 (\bd_1 \cdot \bn )  (\bd_2 \cdot \bn ) \right ] \;  ,
\ee
where
$$
 r \equiv |\br| \; , \qquad \bn \equiv \frac{\br}{r} \; , \qquad
\br \equiv \br_1  - \br_2 \;  .
$$
One often considers the case, where all dipoles are identical and polarized along a unit
vector $\bfe_d$, so that
$$
 \bd_i = d_0 \bfe_d \qquad (d_0 \equiv |\bd_i| ) \;  .
$$
Then potential (\ref{80}) reduces to the form 
\be
\label{81}
 D(\br) = \frac{d_0^2}{r^3} \; \left ( 1 - \cos^2\vartheta \right ) \;  ,
\ee
in which $\vartheta$ is the angle between ${\bf n}$ and the dipole direction,
$$
 \cos\vartheta = \bn \cdot \bfe_d \; .
$$

The dipolar potential, as is easy to check, is not integrable. Therefore the use of
the bare forms, whether (\ref{80}) or (\ref{81}), leads to all those problems described
above. For instance, one confronts the so-called polarization catastrophe 
\cite{Applequist_29,Thole_30}. The necessity of regularizing the dipolar potential has 
been understood long time ago, and several smoothing functions have been suggested for
the regularization at short-range \cite{Thole_30,Burnham_31,Masia_32,Kanjilal_33,Ustunel_34}
as well as at long-range distance \cite{Jonscher_35,Jonscher_36,Jonscher_37,Tarasov_38}.
One of the simplest regularizations, making the potential integrable, results in the 
effective regularized potential 
\be
\label{82}
 D(\br,b,\kappa) = \Theta(r-b) D(\br) e^{-\kappa r} \;  ,
\ee
where $\Theta(r)$ is a unit-step function. This potential is absolutely integrable. And 
the absolutely integrable potential guarantees the existence of the Fourier transform
\be
\label{83}
  D_k(b,\kappa) = \int D(\br,b,\kappa)  e^{-i \bk\cdot\br} d\br \;  ,
\ee
with the inverse transform
$$
  D(\br,b,\kappa) = \frac{1}{V} \sum_k D_k(b,\kappa)  e^{i\bk\cdot\br} \;  .
$$

The Fourier transform (\ref{83}), in the case of the polarized potential (\ref{81}),
gives
\be
\label{84}
  D_k(b,\kappa) = D_k I_k(b,\kappa) \;  .
\ee
This expression is the product of  
\be
\label{85}
 D_k = \frac{4\pi}{3} \; d_0^2 \left ( 3\cos^2\vartheta_k - 1 \right ) \;  ,
\ee
with $\vartheta_k$ being the angle between the vector ${\bf k}$ and the dipole direction,
$$
\cos\vartheta_k = \frac{\bk\cdot\bfe_d}{k} \;  ,
$$
and of the integral
\be
\label{86}
 I_k(b,\kappa) = 9 k b \int_1^\infty \left [ 
\frac{\sin(kbx)}{(kbx)^4} - \frac{\cos(kbx)}{(kbx)^3} - 
\frac{\sin(kbx)}{3(kbx)^2} \right ] e^{-\kappa bx} dx \; .
\ee
The latter, as is seen, depends on two variables $k b$ and $\kappa b$, so that 
it can be presented as 
\be
\label{87}
 I_k(b,\kappa) = J_q(c) \qquad ( q \equiv kb , ~ c \equiv \kappa b ) \;  ,
\ee
with
\be
\label{88}
 J_q(c) = 9q \int_1^\infty \left [ 
\frac{\sin(qx)}{(qx)^4} - \frac{\cos(qx)}{(qx)^3} - 
\frac{\sin(qx)}{3(qx)^2} \right ] e^{-cx} dx \;  .
\ee

Integral (\ref{86}) has the property
\be
\label{89}  
\lim_{b\ra 0} \lim_{\kappa\ra 0} I_k(b,\kappa)  = 1 \;  ,
\ee
because of which
\be
\label{90}
 \lim_{b\ra 0} \lim_{\kappa\ra 0} D_k(b,\kappa)  = D_k \;  .
\ee
This means that in the absence of the regularization, for the interaction potential 
(\ref{81}) we would have the Fourier transform (\ref{85}). However, this transform
is defined neither for $k \ra 0$ nor for $k \ra \infty$, since potential (\ref{81})
is not absolutely integrable. While the Fourier transform (\ref{84}) is well defined
in both these limits,
\be
\label{91}
   \lim_{k\ra 0}  D_k(b,\kappa)  = \lim_{k\ra\infty}  D_k(b,\kappa) = 0 \; .
\ee
For the absolutely integrable potential, it is also admissible to interchange the
limiting operation and integration, so that
\be
\label{92}
 \int D(\br,b,\kappa)\; d\br =  \lim_{k\ra 0}  D_k(b,\kappa) \;  .
\ee
While such an interchange is prohibited for not absolutely integrable potentials.
Really, for the non-regularized potential (\ref{80}), that is not absolutely 
integrable, both sides of the equation similar to Eq. (\ref{92}) would not be 
defined. 

In order to emphasize the problems arising when using not absolutely integrable
potentials, let us take Hamiltonian (\ref{2}), with the dipolar interaction 
potential, and with setting $x \ra {\bf r}$. Employing the regularized potential 
(\ref{82}), for the average energy, in the Hartree-Fock approximation (\ref{4}), 
we have
\be
\label{93}
 \lgl H \rgl = K + \frac{1}{2} \int D(\br - \br',b,\kappa) \left [ \;
\rho(\br) \rho(\br') \pm | \rho(\br,\br') |^2 \right ] \; d\br d\br' \;  ,   
\ee
where
$$
K = \int \lgl \psi^\dgr(\br) K(\br) \psi(\br) \rgl d\br \; , \qquad
\rho(\br) = \lgl \psi^\dgr(\br) \psi(\br) \rgl  \; , \qquad
\rho(\br,\br') = \lgl \psi^\dgr(\br') \psi(\br) \rgl  \; .
$$
For concreteness, let us consider a uniform system, although the same problems 
exist for nonuniform systems, in particular, in the local-density approximation. 
For a uniform system, we get
\be
\label{94}
  \rho(\br) = \rho \; , \qquad \rho(\br,\br') = \rho(\br-\br',0) \; .
\ee
Then energy (\ref{93}) becomes
\be
\label{95}
 \lgl H \rgl = K + \frac{1}{2}\; \rho N  \int D(\br,b,\kappa)\; d\br \pm
\frac{1}{2} \; V \int D(\br,b,\kappa) | \rho(\br,0) |^2 \; d\br \; ,
\ee
with $V$ being the system volume. 

If we would keep the non-regularized potential (\ref{80}) or (\ref{81}) in the
last equation, we would confront divergences. If we use relation (\ref{92}) for
the non-regularized potentials, then energy (\ref{95}) is not defined, since the 
limit (\ref{92}) depends on the type of approaching $k \ra 0$. But then energy  
(\ref{95}) becomes not a scalar, together with other thermodynamic characteristics,
which is, certainly, senseless. Contrary to this, expression (\ref{95}) for the 
regularized potential is well defined.

\section{Iterative procedure for equilibrium systems}

The general iterative procedure, described above, is applicable to any system, 
whether equilibrium or not. It is important to show how it can be employed for 
equilibrium systems. For the latter, the two-point system characteristics, such 
as Green functions and self-energies, depend on the time difference 
$t_{12} \equiv t_1 - t_2$. Therefore one can resort to the Fourier transforms for 
the propagator
\be
\label{96} 
 G(12) = \int G(x_1,x_2,\om) e^{-i\om t_{12}} \frac{d\om}{2\pi}  
\ee
and self-energy
\be
\label{97}
\Sgm(12) = \int \Sgm(x_1,x_2,\om) e^{-i\om t_{12}} \frac{d\om}{2\pi} \; .
\ee
Then the first-order self-energy reads as
$$
\Sgm_1(x_1,x_2,\om) =\dlt(x_1-x_2) \int \Phi(x_1,x_2,\om) \rho(x_3) \; dx_3  \; +
$$
\be
\label{98}
 +\; 
i \int \Phi(x_1,x_2,\om-\om') G(x_1,x_2,\om') \; \frac{d\om'}{2\pi} \;  ,
\ee
in which
\be
\label{99}
 \Phi(x_1,x_2,\om) = \Phi(x_1,x_2) e^{-i\om 0} \;  ,
\ee
with
\be
\label{100}
\Phi(x_1,x_2) = s(x_1,x_2) V(x_1,x_2) 
\ee
and 
\be
\label{101}
 s(x_1,x_2) \equiv \lim_{t_2\ra t_1} s(12) \;  .
\ee
Here, as usual, the expression $\pm \omega 0$ implies $\pm \omega \tau$, with 
$\tau \ra + 0$. 

In the second order, the self-energy becomes
\be
\label{102}
 \Sgm_2(x_1,x_2,\om) = \Sgm_1(x_1,x_2,\om) + \Dlt(x_1,x_2,\om) + \Lbd(x_1,x_2,\om) \; ,
\ee
with the correcting term
$$
\Dlt(x_1,x_2,\om) = \dlt(x_1-x_2) 
\int \Phi(x_1,x_3,\om) [ \; 1 - s(x_1,x_3)\; ] \rho(x_3) \; dx_3 \; +
$$
\be
\label{103}
 + \;
i \int \Phi(x_1,x_2,\om-\om') [ \; 1 - s(x_1,x_2)\; ] G(x_1,x_2,\om') \; \frac{d\om'}{2\pi} 
\ee
and 
$$
\Lbd(x_1,x_2,\om) =
$$
$$
=
 - \int \Phi(x_1,x_3,\om' ) G(x_4,x_3,\om''-\om')  
[\; G(x_1,x_4,\om-\om') G(x_3,x_2,\om'') \Phi(x_4,x_2,\om-\om'') \pm
$$
\be
\label{104}
 \pm
 G(x_1,x_2,\om-\om') G(x_3,x_4,\om'') \Phi(x_2,x_4,-\om')\; ] \; dx_3 dx_4 \; 
 \frac{d\om' d\om''}{(2\pi)^2}  \; .
\ee

To specify these expressions, it is necessary to define the zero-order propagator.
The latter, e.g., can be defined as the expansion
\be
\label{105}
G_0(x_1,x_2,\om) = \sum_k G_k(\om) \psi_k(x_1) \psi_k^*(x_2)
\ee
over the set of orthonormalized wave functions given by the eigenproblem
\be
\label{106}
K(x) \psi_k(x) = E_k \psi_k(x) \;  ,
\ee
where $K(x)$ is the single-particle Hamiltonian entering Eq. (\ref{2}). The index 
$k$ here denotes the set of quantum numbers. It can be momentum for uniform systems
or a set of discrete quantum numbers for finite quantum systems \cite{Birman_40}.
In expansion (\ref{105}), the coefficient function is the Green function in the 
energy representation  
\be
\label{107}
 G_k(\om) = \frac{1\pm n_k}{\om-\om_k + i0} \; \mp \; \frac{n_k}{\om-\om_k-i0} =
P \; \frac{1}{\om-\om_k} - i\pi ( 1 \pm 2n_k) \dlt(\om-\om_k) \;  ,
\ee
with the energy distribution
\be 
\label{108}
 n_k = \frac{1}{\exp(\bt\om_k) \mp 1} \;  ,
\ee
where
\be
\label{109}
 \om_k \equiv E_k - \mu \qquad ( \bt T = 1 ) \;  .
\ee
Here $P$ is the symbol of principal value. 

Then the first-order self-energy (\ref{98}) is
\be
\label{110}
 \Sgm_1(x_1,x_2,\om) = \sum_k n_k [\;\Phi_{kk}(x_1) \dlt(x_1-x_2) 
\pm \Phi(x_1,x_2) \psi_k(x_1) \psi_k^*(x_2)\; ] \;  ,
\ee
where
\be
\label{111}
  \Phi_{kp}(x) \equiv \int \psi_k^*(x') \Phi(x,x') \psi_p(x') \; dx' \; .
\ee
In the second-order self-energy (\ref{102}) for the correcting term, we have 
\be
\label{112}
\Dlt_1(x_1,x_2,\om) = \sum_k n_k \{ B_{kk}(x_1) \dlt(x_1-x_2) 
\pm \Phi(x_1,x_2) [\; 1- s(x_1,x_2) \; ] \psi_k(x_1) \psi_k^*(x_2) \} \;   ,
\ee
with
\be
\label{113}
B_{kp}(x) \equiv 
\int  \psi_k^*(x') \Phi(x,x') [\; 1- s(x,x') \; ] \psi_p(x') \; dx' \;  .
\ee
And the last term in Eq. (\ref{102}), on the complex $\omega$ - plane, has the form
\be
\label{114}
 \Lbd_1(x_1,x_2,\om) = \sum_{ijk} \frac{\Lbd_{ijk}(x_1,x_2)}{\om-\om_{ijk} } \; ,
\ee
in which ${\rm Im}\; \omega \neq 0$,
\be
\label{115}
 \om_{ijk} \equiv \om_i + \om_j - \om_k = E_i + E_j - E_k - \mu \;  ,
\ee
and
$$
\Lbd_{ijk}(x_1,x_2) = \Phi_{ik}(x_1) [\; n_j(n_i - n_k) \pm n_i(1\pm n_k) \; ] 
\times
$$
\be
\label{116}
 \times
\left [\; \Phi_{jk}(x_2)\psi_j(x_1) \psi_k^*(x_2) \pm 
\Phi_{ki}(x_2) \psi_j(x_1) \psi_j^*(x_2) \; \right ] \; .
\ee
The symmetry of $\Phi(x,x')$ has been used.

On the real $\omega$ - axis, we get 
\be
\label{117}
 \Lbd(x_1,x_2,\om) =  \int \Gm(x_1,x_2,\om') \left [ \frac{1\pm n(\om')}{\om-\om'+i0}
\mp \frac{ n(\om')}{\om-\om'- i0} \right ] \frac{d\om'}{2\pi} \; ,
\ee
that can be represented as
\be
\label{118}
 \Lbd(x_1,x_2,\om) = P \int \frac{\Gm(x_1,x_2,\om')}{\om-\om'} \;\frac{d\om'}{2\pi}
 \; - \; \frac{i}{2}\; [ \; 1 \pm 2n(\om) \; ]\; \Gm(x_1,x_2,\om) \; ,
\ee
with the spectral function
\be
\label{119}
 \Gm(x_1,x_2,\om) = i [\; \Lbd(x_1,x_2,\om+i0) - \Lbd(x_1,x_2,\om -i0)\; ] \; .
\ee
The latter, employing Eq. (\ref{114}), becomes
\be
\label{120}
 \Gm(x_1,x_2,\om) = 2\pi \sum_{ijk} \Lbd_{ijk}(x_1,x_2) \dlt(\om-\om_{ijk}) \;  .
\ee
Therefore Eq. (\ref{117}) takes the form
\be
\label{121}
 \Lbd(x_1,x_2,\om) = \sum_{ijk} \Lbd_{ijk}(x_1,x_2) G_{ijk}(\om) \; ,
\ee
with the notations
\be
\label{122}
 G_{ijk}(\om) = \frac{1\pm n_{ijk}}{\om-\om_{ijk} + i0} \; \mp \;
\frac{ n_{ijk}}{\om-\om_{ijk} - i0}
\ee
and
\be
\label{123}
n_{ijk} \equiv \frac{1}{\exp(\bt\om_{ijk})\mp 1} \;   .
\ee
Thus the second-order self-energy contains the real part
\be
\label{124}
{\rm Re}\; \Sgm_2(x_1,x_2,\om) = \Sgm_1(x_1,x_2,\om) +
\Dlt(x_1,x_2,\om)
\ee
and the imaginary part
\be
\label{125}
 {\rm Im}\; \Sgm_2(x_1,x_2,\om) = -\; 
\frac{1}{2} \; [ \; 1 \pm 2n(\om) \; ]\; \Gm_2(x_1,x_2,\om) .
\ee

It is again worth stressing that in all expressions above nowhere we meet the 
bare interaction potential $V(x_1,x_2)$ that would produce divergences, but 
everywhere we have only the smoothed potential $\Phi)x_1,x_2)$.

\section{Iterative calculation of observable quantities}

What one finally needs from any theory is the possibility of calculating observable
quantities. It is, then necessary to show how the suggested iterative procedure 
can be employed for such calculations. One of the most important quantities is 
the internal energy
\be
\label{126}
E = \lgl H \rgl + \mu N \;   .
\ee
Therefore, calculating this quantity is an instructive example demonstrating how
the procedure works.

In terms of Green functions, the Hamiltonian average can be represented as  
\be
\label{127}
 \lgl H \rgl = \pm \; \frac{i}{2} \int \lim_{(21)} \left [
i \; \frac{\prt}{\prt t_1} + K(x_1) - \mu \right ] G(12)\; dx_1 \;  ,
\ee
and the total number of particles as
\be
\label{128}
 N = \pm i \int \lim_{(21)} G(12) \; dx_1 \;  .
\ee
Here, for brevity, we use the notation of the limit
$$
\lim_{(21)} \equiv \lim_{x_2\ra x_1} \; \lim_{t_2\ra t_1+0}  \;  .
$$
In that way, energy (\ref{126}) can be written in the form
\be
\label{129}
 E = \pm \; \frac{i}{2} \int \lim_{(21)} \left [
i \; \frac{\prt}{\prt t_1} + K(x_1) + \mu \right ] G(12) \; dx_1 \;   .
\ee
For an equilibrium system, the latter yields
\be
\label{130}
 E = \pm \; \frac{i}{2} \int e^{+i\om 0} [ \; \om + K(x) + \mu\; ]
G(x,x,\om) \; \frac{d\om}{2\pi} \; dx \; .
\ee
 
We have to substitute into expression (\ref{130}) the approximate Green functions
obtained by means of the above iterative procedure. In the process of these
calculations, there arise the following delicate point. In the integral over
frequency $\omega$, there appear the products of the functions $G_k(\omega)$
defined in Eq. (\ref{107}), including the products of the Green functions with
coinciding poles, such as $G_k^n(\omega)$, where $n = 1, 2, \ldots$. Direct
integration over such expressions $G_k^n(\omega)$ results in divergent integrals. 
This is caused by the fact that Green functions are distributions 
(generalized functions), which are not well defined for the products with coinciding 
poles \cite{Bogolubov_48,Yukalov_49}. Such products require additional definition. 
The method of dealing with the integrals over the products of Green functions with 
coinciding poles, used in the present paper, is described in Appendix A.   

The initial zero approximation for the energy corresponds to the use of the Green
function (\ref{105}), which gives
\be
\label{131}
 E^{(0)} = \sum_k n_k E_k \;  .
\ee
The first-order propagator reads as
\be
\label{132}
 G_1(x_1,x_2,\om) =  G_0(x_1,x_2,\om) + 
\sum_{kp} G_k(\om) G_p(\om) M_{kp} \psi_k(x_1) \psi_p^*(x_2) \; ,
\ee
where
\be
\label{133}
M_{kp} = \sum_m n_m ( \Phi_{kmmp} \pm \Phi_{kmpm} ) \;   ,
\ee
with the matrix elements
\be
\label{134}
 \Phi_{mkpn} \equiv \int \psi_m^*(x) \Phi_{kp}(x) \psi_n(x) \; dx \;  .
\ee
Then the first-order energy becomes
\be
\label{135}
E^{(1)} = E^{(0)} + 
\frac{1}{2} \sum_k n_k M_{kk} [ \; 1 - 2\bt ( 1 \pm n_k) E_k \; ] \;   .
\ee

The second-order propagator takes the form
$$
G_2(x_1,x_2,\om) =  G_1(x_1,x_2,\om) +\Dlt G(x_1,x_2,\om) +
$$
\be
\label{136}
  +
\sum_{mn} \; \sum_{ijk} G_m(\om) G_n(\om) G_{kji}(\om) \Lbd_{ijk}^{mn} \psi_m(x_1)\psi_n^*(x_2) \;.
\ee
Here the correcting term is
\be
\label{137}
 \Dlt G(x_1,x_2,\om) = \sum_{mn}  G_m(\om) G_n(\om) \Dlt_{mn}\psi_m(x_1)\psi_n^*(x_2) \;   ,
\ee
in which
\be
\label{138}
   \Dlt_{mn} = \sum_k n_k ( B_{mkkn} \pm B_{mknk} ) 
\ee
and
\be
\label{139}
 B_{mkpn} \equiv \int  \psi_m^*(x) B_{kp}(x) \psi_n(x)\; dx \; ,
\ee
with the matrix elements $B_{kp}(x)$ being defined in Eq. (\ref{113}). The last 
term in propagator (\ref{136}) contains
\be
\label{140}
 \Lbd_{ijk}^{mn} = [\; n_j( n_i - n_k) \pm n_i ( 1 \pm n_k) \; ]
\Phi_{mikj} ( \Phi_{kjin}  \pm \Phi_{jkin} ) \;  .
\ee

This propagator yields the second-order approximation for the energy
\be
\label{141}
 E^{(2)} = E^{(1)}  +\Dlt E + \frac{1}{2} \sum_n \; \sum_{ijk} \Lbd_{ijk}^{nn} \left (
E_n C_n^{ijk} + D_n^{ijk} \right ) \; ,
\ee
with the correcting term
\be
\label{142}
\Dlt E = \frac{1}{2} \sum_k \Dlt_{kk} n_k [ \; 1 - 2\bt E_k ( 1 \pm n_k) \; ] \;   .
\ee
Here the notations 
\be
\label{143}
C_n^{ijk} \equiv \frac{I_{nn} - I_n^{ijk} }{\om_n - \om_{ijk}}
\ee
and 
\be
\label{144}
D_n^{ijk} \equiv \frac{\om_n I_{nn} - \om_{ijk} I_n^{ijk} }{\om_n - \om_{ijk}}
\ee
are used, in which
\be
\label{145}
I_{kk} = - \bt n_k(1 \pm n_k) \; , \qquad
 I_p^{ijk} = \frac{n_p-n_{ijk}}{\om_p-\om_{ijk}} \;  .
\ee
Also, notations (\ref{115}) and (\ref{123}) are employed.

\section{Illustration of smallness of correcting terms}

As is seen from the above expressions, the correcting terms for the internal 
energy contain the matrix elements
$$
B_{kppk} = \int | \psi_k(x) |^2 | \psi_p(x') |^2 \Phi(x,x') [ \; 1 - s(x,x') \; ]\;
dx dx'
$$
and
$$
B_{kpkp} = \int \psi_k^*(x)  \psi_p^*(x')  \Phi(x,x') [ \; 1 - s(x,x') \; ]
\psi_k(x')  \psi_p(x)\; dx dx'  \; ,
$$
which should be compared with the matrix elements
$$
\Phi_{kppk} = \int | \psi_k(x) |^2 | \psi_p(x') |^2 \Phi(x,x')\;  dx dx'
$$
and
$$
\Phi_{kpkp} = \int \psi_k^*(x)  \psi_p^*(x')  \Phi(x,x') 
\psi_k(x')  \psi_p(x)\; dx dx'  \;   .
$$

In order to show that the correcting terms are usually much smaller than
the main terms, let us consider a uniform system, for which the natural orbitals
are the plane waves 
$$
 \psi_k(\br) = \frac{1}{\sqrt{V}} \; e^{i\bk\cdot\br} \;  .
$$
The role of the variable $x$ is played by the spatial variable ${\bf r}$. The bare 
interaction potential is $V({\bf r} - {\bf r}')$ and the smoothing function is
$s({\bf r} - {\bf r}')$, respectively the smoothed effective potential also depends
on the difference ${\bf r} - {\bf r}'$, being $\Phi{\bf r} - {\bf r}')$. Moreover,
the standard situation is when the interaction potentials depend on the absolute
value $|{\bf r} - {\bf r}'|$, which we shall keep in mind, so that 
$\Phi({\bf r}) = \Phi(r)$, where $r \equiv |{\bf r}|$.

Then the matrix element $\Phi_{kppk}$ reduces to
\be
\label{146}
 \Phi_0 = 4\pi \int_0^\infty \Phi(\br) r^2 \; dr  
\ee
and the matrix element $B_{kppk}$, to
\be
\label{147}
 B_0 = 4\pi \int_0^\infty \Phi(\br) [ \; 1 - s(r) \; ] r^2 \; dr \;  .
\ee
The main contribution from the exchange elements $\Phi_{kpkp}$ and $B_{kpkp}$ is 
usually smaller than that from the direct elements $\Phi_{kppk}$ and $B_{kppk}$,
respectively, so that it is sufficient to compare the values of expressions (\ref{146}) 
and (\ref{147}).    

For illustration, let us consider the Lennard-Jones potential (\ref{78}), with
the smoothing function (\ref{79}). Hence the smoothed effective potential is
\be
\label{148}
 \Phi(\br) = 4\ep \left [ \left ( \frac{\sgm}{r} \right )^{12} - 
\left ( \frac{\sgm}{r} \right )^6 \right ] \; \exp \left \{ - b_0 
\left ( \frac{\sgm}{r} \right )^5 \right \} \; .
\ee
Then for expression (\ref{146}), we find
\be
\label{149}
 \frac{\Phi_0}{16\pi\ep \sgm^3} = \frac{1}{5} \left [
\frac{\Gm(9/5)}{b_0^{9/5} } \; - \;  \frac{\Gm(3/5)}{b_0^{3/5} } \right ] \; ,
\ee
while for expression (\ref{147}), 
\be
\label{150}   
 \frac{B_0}{16\pi\ep \sgm^3} = \frac{1}{20} \left [ 
\left ( 4-2^{1/5} \right ) \; \frac{\Gm(9/5)}{b_0^{9/5} }  - 
\left ( 4-2^{7/5} \right ) \;  \frac{\Gm(3/5)}{b_0^{3/5} } \right ] \;,
\ee
where the relation $\Gamma(x-1) = \Gamma(x)/(x-1)$ is used.
 
Taking, for concreteness, the value $b_0 = 1.86$ corresponding to $^4$He, we
obtain
$$
 \frac{B_0}{\Phi_0} \sim 0.1 \;  .
$$
This demonstrates that the correcting terms are an order smaller than the main 
terms, hence, to a good approximation, the former can be omitted.

\section{Rules for defining smoothing functions}

The general iterative procedure is formulated with a necessary requirement that 
smoothing functions, regularizing interaction potentials, be such that the 
regularized effective potentials be integrable, which can be written as the
condition
\be
\label{151}
\left | \int V(x_1,x_2) s(x_1,x_2) \; dx_2 \right | < \infty \;   .
\ee
This implies that, when the bare interaction potential diverges, this divergence has
to be compensated by the tendency of the smoothing function to zero, hence
\be
\label{152}
 s(x_1,x_2) \ra 0 \; , \qquad V(x_1,x_2) \ra \infty \;  .
\ee
From the other side, if the bare potential becomes small, there is not need in the  
regularization, so that the smoothing function should tend to one:
\be
\label{153}
  s(x_1,x_2) \ra 1 \; , \qquad V(x_1,x_2) \ra 0 \;  .
\ee
These are the general conditions imposed on any smoothing function, for which the
iterative procedure has sense. 

It is straightforward to notice that there is a physical quantity satisfying these 
conditions - this is the pair correlation function
\be
\label{154}
 g(x_1,x_2) = \frac{\lgl\; \hat n(x_1) \hat n(x_2)\; \rgl }{\rho(x_1)\rho(x_2)} \; ,
\ee
with the density operator
$$
 \hat n(x) \equiv \psi^\dgr(x) \psi(x) \;  .
$$
Therefore, the smoothing function can be associated with the pair correlation function
taken in some approximation.

A simple way of constructing a smoothing function $s(x_1 - x_2)$, as a correlation 
function, is by defining it through the wave function $\chi(x)$ of the relative motion 
of two scattering particles,
$$
 s(x) \propto |\; \chi(x) \; |^2 \;  ,
$$  
keeping in mind the boundary conditions (\ref{152}) and (\ref{153}). For example, if 
the bare potential diverges at short distance as
$$
 V(r) \simeq 4\ep \left ( \frac{\sgm}{r} \right )^n \qquad ( r \ra 0 ) \;  ,
$$
then we find
\be
\label{155}
s(r) = \exp \left \{ - b_0 \left ( \frac{\sgm}{r} \right )^{(n-2)/2} \right \} \; ,
\ee
where
$$
b_0 \equiv \frac{8}{(n-2)\Lbd} \qquad 
\left ( \Lbd \equiv \frac{1}{\sqrt{m\ep \sgm^2}} \right ) \;  . 
$$
Substituting here $n = 12$, we get the smoothing function used above for the 
Lennard-Jones interaction potential.

\section{Extrapolation to large coupling parameters}

The correlated iterative procedure, described in the previous sections, makes it
possible to find successive approximations for observable quantities, without 
confronting divergences at any step, despite that the bare interaction potential
can be nonintegrable. As follows from the structure of the terms arising in
this iterative procedure, the difference between the iterative cases, starting 
with either a bare nonintegrable interaction potential $V(12)$ or with an 
integrable smoothed potential $\Phi(12)$, is in the appearance of correcting terms 
containing the expression $1 - s(12)$ in front of the smoothed potential $\Phi(12)$. 
Estimating the correcting terms, we have shown that they are small, as compared to 
the main terms, when the smoothing function is chosen as an approximate pair 
correlation function. The smallness becomes evident, even without numerical 
calculations, when the particle interactions are small, since when $\Phi(12) \ra 0$, 
then $s(12) \ra 1$, hence the product $\Phi(12) [1-s(12)]$ quickly tends to zero. 

Thus it is possible to find the successive terms of the iterative procedure. But 
the following question remains: Can we get a convergent series of such terms? 

Suppose that it is admissible to replace the bare nonintegrable potential by an 
integrable smoothed potential, as has been discussed above. But the iterative 
procedure yields the approximations having the structure of series in powers of 
the smoothed potential. 

It is worth recalling that series in powers of interactions practically always 
are divergent. This is well known for the standard perturbation theory with Green
functions, even when the interaction potentials are perfectly integrable \cite{Kadanoff_9}. 
Moreover, even the simplest example of an anharmonic oscillator, being treated 
with the standard Rayleigh-Schr\"{o}dinger perturbation theory, results in series 
that are divergent for any finite value of the coupling parameter. Perturbative
or iterative series are well known to be asymptotic, having sense only for 
asymptotically small coupling parameters.    

Then the general and natural question is: Having a series in powers of a weak 
coupling parameter, is it feasible to extrapolate it to large values of the coupling 
parameter? The answer is "yes", however, such an extrapolation requires involving 
additional methods based on self-similar approximation theory \cite{Yukalov_50,Yukalov_51}.

To be more precise, let us define the dimensionless {\it coupling parameter} as 
the ratio of the effective interaction strength to effective kinetic-energy 
strength,
\be
\label{156}
 g \equiv \frac{m}{4\pi a} \int \Phi(\br) \; d\br \;  ,
\ee
where $a$ is mean-interparticle distance. In the case of a spherically symmetric
potential, this reduces to 
\be
\label{157}
  g \equiv \frac{m}{a} \int_0^\infty \Phi(\br) r^2 \; d\br \;  .
\ee

Suppose we are calculating an observable quantity that is the statistical average
of a self-adjoint operator, for instance, this can be the internal energy, as is 
considered above. Let us denote this observable as $f(g)$, which is a function of 
the coupling parameter $g$. The $k$-th order series in powers of the coupling 
parameter has the general form
\be
\label{158}
f_k(g) = f_0(g) \left ( 1 + \sum_{m=1}^k a_m g^m \right ) \;   ,
\ee
where $f_0(g)$ is the known initial approximation. For realistic problems, such 
series are practically always divergent for any finite value of $g$. Moreover, 
for the majority of interesting problems, one is able to calculate only the 
second-order approximation
\be
\label{159}
 f_2(g) = f_0(g) \left ( 1 +  a_1 g + a_2 g^2 \right ) \;  ,
\ee
since the higher-order approximations become untreatably cumbersome. 

We know that, if the coupling parameter is not too large, the described iterative 
procedure, using an effective smoothed potential, is perfectly admissible, since 
the correcting terms, as is shown above, are small. This is in agreement with the
studies \cite{Kalos_52,Giorgini_53} showing that, under weak interactions, the 
results are weakly dependent on the shape of the used potential. But the question 
remains: How the obtained result can be extrapolated to large values of the coupling 
parameter?

The effective extrapolation from small $g$ to large $g$ can be done involving the 
self-similar approximation theory \cite{Yukalov_50,Yukalov_51} in the frame of 
self-similar factor approximants \cite{Yukalov_54,Gluzman_55,Yukalov_56}. We shall 
not go into the details of the self-similar approximation theory, whose thorough 
exposition has been done in the published papers 
\cite{Yukalov_50,Yukalov_51,Yukalov_54,Gluzman_55,Yukalov_56}, but let us just apply 
it to the second-order expansion (\ref{159}). Then the second-order factor 
approximant, extrapolating the weak-coupling expansion (\ref{159}) to finite values 
of $g$, reads as
\be
\label{160}
  f_2^*(g) = f_0(g) ( 1 + A g  )^n \; ,
\ee
with the parameters
$$
 A = \frac{a_1^2 - a_2}{a_1} \; , \qquad n = \frac{a_1^2}{a_1^2-a_2} \;  .
$$

As an example of a nonintegrable potential, let us take the hard-core potential
$V(r)$ that is zero for $r > \sigma$ and becomes infinite for $r \leq \sigma$. It
is known \cite{Kalos_52,Giorgini_53} that in the low-energy region this potential 
can be replaced by the pseudopotential
\be
\label{161}
  \Phi(\br) = 4\pi \; \frac{a_s}{m} \; \dlt(\br) \; ,
\ee
in which $a_s$ is scattering length equal to the diameter $\sigma$ of the hard core. 
With this potential (\ref{161}), the coupling parameter (\ref{157}) becomes
\be
\label{162}
 g = \frac{a_s}{a} = \rho^{1/3} a_s \;  ,
\ee
where $\rho$ is average density, such that $\rho a^3 = 1$.   
     
As an example, let us consider the ground-state energy of a dilute Bose system, 
introducing the dimensionless energy
\be
\label{163}
 E_0 \equiv 2m a_s^2 \; \frac{E}{N} \qquad ( T = 0 ) \;  .
\ee
For asymptotically weak coupling $g \ra 0$, the ground-state energy of a uniform 
system is found \cite{Lee_57,Lee_58,Lee_59,Wu_60} to be
\be
\label{164}
 E_{LHY}(g) = 4\pi g^3 
\left ( 1 + \frac{128}{15\sqrt{\pi}} \; g^{3/2} \right ) \;  .
\ee

In the asymptotic region, where $g \ra 0$, the pseudopotential (\ref{161}) is known  
\cite{Kalos_52,Giorgini_53} to well describe the system with hard-core interactions.
But can the use of such an effective potential be somehow extrapolated to finite 
values of the coupling parameter? Actually, dealing with a uniform system, one needs 
to consider only the region $g \in [0,0.6]$, since at the critical value $g_c = 0.65$, 
the system crystallizes, becoming nonuniform \cite{Rossi_43}.   

To realize the extrapolation by means of self-similar factor approximants, we consider 
a uniform Bose system at zero temperature with the effective interaction potential 
(\ref{161}), calculate the ground-state energy, with the separated factor 
$f_0(g) = 4 \pi g^3$, in the second order \cite{Yukalov_42}, with respect to 
$z \equiv g^{3/2}$, and employ the self-similar approximation theory, which yields
\be
\label{165}
 E_0(g) = 4\pi g^3 \left ( 1 + 2.93379 g^{3/2} \right )^{1.64103} \;  .
\ee
This formula exactly reproduces the Lee-Huang-Yang expression (\ref{164}) for small
$g$ and practically coincides with the Monte Carlo simulations \cite{Rossi_43} for 
all coupling parameters in the region $0 \leq g \leq 0.6$, where the system can be 
treated as uniform. 

This example demonstrates that the use of an effective integrable interaction 
potential, complimented by self-similar approximation theory, can accurately 
reproduce the properties of systems with nonintegrable interaction potentials,
such as the hard-core potential, in a wide range of coupling parameters, hence 
extrapolating the series for asymptotically small coupling parameters to their 
finite values.

\section{Conclusion}

In the paper, statistical systems are considered composed of atoms interacting 
trough nonintegrable interaction potentials. The treatment of such potentials, as  
is well known, confronts several problems, such as the impossibility of using the 
standard mean-field approximations, for instance, Hartree, Hartree-Fock, or
Hartree-Fock-Bogolubov approximations, the impossibility of introducing coherent 
states, the difficulty in breaking the global gauge symmetry, required for 
describing Bose-Einstein condensed and superfluid systems, and the absence of 
correctly defined Fourier transforms that are needed for characterizing uniform
systems as well as nonuniform systems in the local-density approximation.  

An efficient iterative procedure for describing such systems is developed, starting 
from a correlated mean-field approximation, with a regularized interaction 
potential, allowing for a systematic derivation of higher orders, and meeting no 
problems arising when employing non-regularized potentials. 

The admissibility of using, instead of bare interaction potentials, leading to 
divergences, some kind of pseudopotentials is known for many quantum systems
in a mean-field approximation \cite{Birman_40}. The principal result of the 
present paper is in proving that it is possible to develop a regular iterative 
procedure for deriving higher-order approximations above the mean-field one
and meeting no divergences at any step. It is also shown that the iterative 
procedure, based on the nonintegrable bare interaction potential, can be 
reorganized in such a way, where the first-order approximation coincides with 
the mean-field approximation with a regularized potential and the higher orders
are close to those that correspond to the standard iterative procedure based on 
the regularized potential. This justifies the use of the regularized potentials
not only in the mean-field approximation, but in the higher orders of the 
iterative procedure as well.        
   
The iterative procedure is specified for equilibrium systems and its application
is illustrated by the calculation of observable quantities, such as internal energy. 
For the case of the Lennard-Jones interaction potential, it is demonstrated that 
the correcting terms, distinguishing the iterative procedures starting with a 
nonintegrable bare potential and with an integrable effective potential, are small. 

Complimenting the iterative procedure by self-similar approximation theory,
it is possible to extrapolate the results, derived for weak coupling, to large
values of coupling parameters. For instance, the obtained formula for the 
ground-state energy of a uniform Bose system practically coincides with the results
of accurate Monte Carlo simulations in the whole region of the coupling parameter,
where the system is uniform, and yields the expression exactly reproducing the
Lee-Huang-Yang limit for weak coupling.

\vskip 2mm

{\bf Acknowledgement}. Financial support form the RFBR (grant $\# $14-02-00723)
is acknowledged. I am grateful for discussions to E.P. Yukalova.

\section*{Appendix A}

In the process of calculation of observable quantities, one meets the integrals
over the products of Green functions, which should be treated with caution, since,
under coinciding poles, such integrals diverge. One meets the integrals of the type
$$
I_{mn} = \pm i \int_{-\infty}^\infty e^{+i\om 0} G_m(\om) G_n(\om) \; \frac{d\om}{2\pi} \; ,
\qquad
J_{mn} = \pm i \int_{-\infty}^\infty e^{+i\om 0} \om G_m(\om) G_n(\om) \; \frac{d\om}{2\pi} \;,
$$
$$
I_{ijk} = \pm i \int_{-\infty}^\infty e^{+i\om 0} G_i(\om) G_j(\om) G_k(\om) \; 
\frac{d\om}{2\pi} \; ,
\qquad
J_{ijk} = \pm i \int_{-\infty}^\infty e^{+i\om 0} \om G_i(\om) G_j(\om) G_k(\om) \; 
\frac{d\om}{2\pi} \;   .
$$
In the expressions for observable quantities, these integrals often enter having
the diagonal form with respect to their indices, which implies the coinciding poles
of the Green functions in the integrands. However, for coinciding poles, the integrals 
diverge, since the products of distributions with coinciding poles are not well defined. 
This problem can be treated in two ways.

One possibility is to consider, under integration, the poles as being different, 
which gives
$$
I_{kp} = \frac{n_k-n_p}{\om_k-\om_p} \; , \qquad
J_{kp} = \frac{\om_k n_k- \om_p n_p}{\om_k-\om_p} \; , 
$$
$$
I_{ijk} = R_{ijk} + R_{jki} + R_{kij} \; , \qquad
J_{ijk} = \om_i R_{ijk} + \om_j R_{jki} + \om_k R_{kij} \; ,
$$
where
$$
R_{ijk} = 
\frac{n_i(1\pm n_j)(1\pm n_k) \pm (1\pm n_i) n_j n_k}{(\om_i-\om_j)(\om_i-\om_k)} \; .
$$
And then to accomplish the limiting procedure to equal indices, which results in
the following limits, for two coinciding poles,
$$
I_{kk} \equiv \lim_{p\ra k} I_{pk} = - \bt n_k ( 1 \pm n_k) \; ,
$$
$$
J_{kk} \equiv \lim_{p\ra k} J_{pk} =  n_k [\; 1 - \bt \om_k(1 \pm n_k) \; ] \; ,
$$
$$
I_{njn} \equiv \lim_{m\ra n} I_{mjn} = \frac{I_{nn}-I_{nj}}{\om_n-\om_j} \; , \qquad  
J_{njn} \equiv \lim_{m\ra n} J_{mjn} = \frac{\om_n I_{nn}-\om_j I_{nj}}{\om_n-\om_j} \;,
$$
and for three coinciding poles,
$$
I_{kkk} \equiv \lim_{p\ra k} I_{kpk} = 
\frac{1}{2} \; \bt^2 n_k ( 1 \pm n_k) ( 1 \pm 2n_k) \; ,
$$
$$
J_{kkk} \equiv \lim_{p\ra k} J_{kpk} =  -\bt n_k ( 1 \pm n_k) \left [ 1 - \;
\frac{1}{2}\; \bt \om_k ( 1 \pm n_k ) \right ] \; .
$$

The other, faster, way is to define the product of $m$ Green functions with 
coinciding poles as
$$
 G_k^m(\om) \equiv \frac{1\pm n_k}{(\om-\om_k+i0)^m} \; \mp \; 
\frac{n_k}{(\om-\om_k-i0)^m}\;  .
$$ 
Then employing the integration
$$
 \pm i \int_{-\infty}^\infty e^{+i\om 0} f(\om) G_k^m(\om) \; \frac{d\om}{2\pi} =
\frac{1}{(m-1)!} \; \frac{d^{m-1}}{d\om_k^{m-1}} \; [\; f(\om_k) n_k  \; ] \; ,
$$
and using the derivatives
$$
 \frac{d n_k}{d\om_k} = -\bt n_k ( 1 \pm n_k) \; , \qquad
\frac{d^2 n_k}{d\om_k^2} = \bt^2 n_k ( 1 \pm n_k) ( 1 \pm 2n_k) \;  ,
$$
one comes to the same expressions as in the first way.

\end{document}